\documentclass[aps,prl,twocolumn,showpacs]{revtex4}

\usepackage{mathptm}    
\usepackage{dcolumn}                    
\usepackage{bm}                         
\usepackage{graphicx}

\newcommand {\ba} {\begin{eqnarray}}
\newcommand {\ea} {\end{eqnarray}}

\usepackage{times}

\begin{document}

\title{Impurity suppression of the critical temperature in the iron-based superconductors}

\author{Yunkyu Bang}
\email[]{ykbang@chonnam.ac.kr} \affiliation{Department of Physics,
Chonnam National University, Kwangju 500-757, Republic of Korea,
and Asia Pacific Center for Theoretical Physics, Pohang 790-784,
Republic of Korea}

\author{Han-Yong Choi}
\email[]{hychoi@skku.ac.kr} \affiliation{Department of Physics and
Institute for Basic Science Research, SungKyunKwan University,
Suwon 440-746, Republic of Korea}

\author{Hyekyung Won}
\email[]{hkwon@hallym.ac.kr} \affiliation{Department of Physics,
Hallym University, Chuncheon 200-702, Republic of Korea}

\begin{abstract}
We study the impurity suppression of the critical temperature
$T_c$ of the FeAs superconductors theoretically based on the the
$\pm$s-wave pairing state of a two band model. The effects of
non-magnetic and magnetic impurities are studied with the
$\mathcal{T}$-matrix approximation, which can continuously treat
impurity scattering from weak to strong coupling limit.
We found that both magnetic and non-magnetic impurities suppress
$T_c$ with a rate that is practically indistinguishable from the
standard d-wave case despite a possibly large difference of the
positive and negative s-wave order parameter (OP) magnitudes. This
is because the density of states enters together with the OP
magnitude for the scattering process.
\end{abstract}


\date{\today}
\maketitle

{\it Introduction --} The alluring prospect of opening a key
window to understanding the mechanism of high temperature
superconductivity (SC) has attracted fierce research activities in
the iron based pnictides \cite{YKamihara,PhysToday08}. The first
step towards this goal is to establish the pairing symmetry of the
FeAs superconductors. Many ideas have been put forward to
understand the seemingly conflicting experimental observations on
the FeAs materials with regard to the pairing symmetry. Among
them, particularly appealing is the sign reversing pairing state
proposed by Mazin and coworkers \cite{Mazin08}. It is the ground
state of a two band superconductivity where both pairing order
parameters on the two bands have full gaps while acquiring the
$\pi$ phase shift between them, which is referred to as the
$\pm$s-wave pairing state \cite{Choi08}. It was noticed early on
that there is this type of solution to a multi-band BCS gap
equation \cite{Suhl59,Rice91}, and it is quite exciting that it
seems to be actually realized in the pnictide superconductors. A
repulsive interband interaction is turned to induce pairing by
generating the sign reversal between the two pairing order
parameters. The $\pm$s-wave state seems to be able to explain most
of the experimental observations indicating the full gap behavior
\cite{pene} as well as a gapless behavior
\cite{T1,Bang08b,recent}.

In this paper, we wish to show that the relative phase of $\pi$
shows up in the impurity suppression of the critical temperature
$T_c$ in an interesting way. We employ the $\mathcal{T}$-matrix
approximation in the weak coupling two band BCS theory. In the
previous paper, using the same theoretical method, we reported
that the impurity effects on the $\pm$s-wave state can introduce
an unusual behavior in NMR $1/T_1$ relaxation rate \cite{Bang08b}.
Therefore, it would be interesting to study the effect of impurity
on the $T_c$ suppression in this unconventional pairing state. The
$T_c$ suppression by non-magnetic impurity is, as might be
expected from the sign changing gap nature of the $\pm$s-wave
state, in between the $s$-wave and $d$-wave pairing states.
Unexpected, however, is that it is indistinguishably close to the
standard d-wave case despite a large difference of the positive
and negative OP magnitudes. We also found that magnetic impurities
are more efficient pair breakers than non-magnetic impurities in
the $\pm$s-wave pairing state and therefore magnetic impurities
yield a faster $T_c$ suppression rate in the $\pm$s-wave pairing
state than in the $d$-wave pairing state although it is a marginal
difference with realistic parameters.

{\it Formalism --} We study a two band model for the FeAs
superconductors (SC). The details were presented in the reference
\cite{Bang08}. Assuming two SC order parameters, $\Delta_h$ and
$\Delta_e$ on each band, the two coupled gap equations are written
as
 \ba
\Delta_h (k) &=& - \sum_{k' } \left[V_{hh} (k,k') \chi_h (k') +
V_{he} (k,k') \chi_e (k') \right], \nonumber \\
 \Delta_e (k) &=& -
\sum_{k' } \left[V_{eh} (k,k') \chi_h (k') + V_{ee}(k,k')  \chi_e
(k') \right],
 \ea
where $V_{h,e} (k,k')$ is the phenomenological pairing interaction
originating from the antiferromagnetic (AFM) correlation. The
above gap equation permits two solutions. When the inter-band
pairing interaction $V_{he}=V_{eh}$ are repulsive and dominant
over the intra-band interactions, the state where $\Delta_h$ and
$\Delta_e$ have the relative phase of $\pi$, referred to as
$\pm$s-wave pairing state, is the ground state. The pair
susceptibility is given by
\begin{eqnarray}
\chi_{h,e}(k) &=& T \sum _n N(0)_{h,e} \int _{-\omega_{AFM}}
^{\omega_{AFM}} d \xi \frac{ \tilde{\Delta}_{h,e}(k) } {
\tilde{\omega}_n^2 +\xi^2 + \tilde{\Delta}_{h,e} ^2 (k)},
\end{eqnarray}
where $N(0)_{h,e} $ are the DOS of the hole and electron bands,
respectively, and $\omega_{AFM}$ is the cutoff energy of the
pairing potential $V(q)$.

The impurity effects are included within the $\mathcal{T}$-matrix
approximation as
 \ba
\tilde{\omega}_n =\omega_n + \Sigma^0 _h(\omega_n) + \Sigma^0
_e(\omega_n), \nonumber \\
\tilde{\Delta}_{h,e} = \Delta_{h,e} + \Sigma^1 _{h} (\omega_n) +
\Sigma^1 _{e} (\omega_n), \nonumber \\
\Sigma_{h,e} ^{0,1} (\omega_n)  = \Gamma \cdot T^{0,1} _{h,e}
(\omega_n), ~~ \Gamma= \frac{n_{imp}}{\pi N_{tot}},
 \ea
where $\omega_n= T \pi (2n +1)$ is the Matsubara frequency,
$n_{imp}$ the impurity concentration, and $N_{tot}=N_h(0) +N_e(0)$
is the total DOS. The $\mathcal{T}$-matrices $\mathcal{T}^{0,1}$
are the Pauli matrices $\tau^{0,1}$ components in the Nambu space.
The impurity induced self-energies are calculated with the
$\mathcal{T}$-matrix generalized to a two band superconductivity
as \cite{Bang08b}, \ba
\mathcal{T}^{i} _{a} (\omega_n) &=& \frac{G^{i} _{a} (\omega_n)}{D} ~~~~~(i=0,1; ~~a=h,e), \\
 D &=& c^2 +[G^0 _h + G^0 _e]^2 + [G^1 _h + G^1 _e]^2,\\
G^0 _a (\omega_n) &=& \frac{N_a}{N_{tot}} \left\langle
\frac{\tilde{\omega}_n}
{\sqrt{\tilde{\omega}_n^2 + \tilde{\Delta}_{a} ^2 (k) }} \right\rangle,\\
G^1 _a (\omega_n) &=& \frac{N_a}{N_{tot}} \left\langle
\frac{\tilde{\Delta}_{a}} {\sqrt{\tilde{\omega}_n^2 +
\tilde{\Delta}_{a} ^2 (k) }}  \right\rangle,
\ea where $c=\cot
\delta_0$ is a convenient measure of scattering strength, with
$c=0$ for the unitary limit and $c > 1$ for the Born limit
scattering. $\langle ...\rangle$ denotes the Fermi surface
average.

Because we are interested in determining $T_c$, we take $T
\rightarrow T_c$ limit and linearize the gap equation with respect
to the order parameters. We obtain
 \ba
\tilde{\omega}_n
&=& \omega_n (1+ \eta_{\omega}), \\
\tilde{\Delta}_{h,e} &=& \Delta_{h,e} (1+\delta_{h,e}),
 \ea
where
 \ba
\eta_{\omega} &=& \frac{\Gamma}{1+c^2} \frac{1}{|\omega_n|},  \\
\delta_{h,e} &=& \frac{\Gamma}{1+c^2} \frac{1}{|\tilde{\omega}_n|}
\frac{[\tilde{N_h}\Delta_{h}+ \tilde{N_e}\Delta_{e}]}
{\Delta_{h,e}},
 \ea
with $\tilde{N_a}=N_a /N_{tot}$. The pair susceptibility can be
written as
 \ba
 \label{chi_he}
\chi_{h,e}(k) = \pi T \sum _n N(0)_{h,e} \frac{ \Delta_{h,e}(k)
(1+\delta_{h,e}) } {|\omega _n (1+ \eta_{\omega})|}.
 \ea
It is immediately clear that $\eta_{\omega}=\delta_{a}$ for a
single band $s$-wave gap state and there is no renormalization of
the pair susceptibility $\chi_{a}(k)$ with the impurity
scattering. This is just the Anderson theorem of $T_c$ for the
$s$-wave SC. In our two band case, it is more complicated to draw
any simple conclusion. In particular, the signs of $\delta_h$ and
$\delta_e$ are opposite because of the opposite signs of
$\Delta_{h,e}$.

Before we show the numerical results we can analyze a simpler
case. The main pairing process in the $\pm$s-wave pairing state is
the inter-band interaction so that we keep only $V_{he}=V_{eh}$
interactions in the gap Eqs.\ (1) and (2), and use Eq.\
(\ref{chi_he}) to obtain
\begin{eqnarray}
 \label{del_spi}
\Delta_h &=& \pi^2 T^2 \sum _n   \sum _m \lambda_{eff} ^2 \frac{
(1+\delta_{e}) (1+\delta_{h}) } {|\omega _n (1+ \eta_{\omega})|
|\omega _m (1+ \eta_{\omega})|}  \Delta_{h},
\end{eqnarray}
where $\lambda_{eff}  = \sqrt{N_h N_e V_{he} V_{eh}}$ is the
effective dimensionless coupling constant. This equation can be
compared with the similarly reduced gap equation without
impurities as
\begin{eqnarray}
 \label{del_s}
\Delta_h &=& \pi^2 T^2 \sum _n   \sum _m \lambda_{eff} ^2 \frac{ 1
} {|\omega _n | |\omega _m| } \Delta_{h} ,
\end{eqnarray}
which yields the standard single band $s$-wave result with $T_c ^0
\approx 1.14 \omega_{D} \exp {(-1/\lambda_{eff})}$. Eq.\
(\ref{del_spi}) would yield definitely smaller $T_c$ than $T_c ^0$
because $\delta_{a}$ is smaller in magnitude than $\eta_{\omega}$.
When both $\delta_{a}$ are set to zeros we obtain another reduced
gap equation as
\begin{eqnarray}
\Delta_h &=& \pi^2 T^2 \sum _n   \sum _m \lambda_{eff} ^2 \frac{ 1
} {|\omega _n (1+ \eta_{\omega})| |\omega _m (1+ \eta_{\omega})|}
\Delta_{h}
\end{eqnarray}
which is just the case that we would obtain for a double $d$-wave
pairing state \cite{Bang08} where the anomalous self-energy
corrections ($\delta_{a}$) are absent because of the sign-changing
OP with equal sizes. Our case of Eq.\ (\ref{del_spi}) is not
straightforward. If both $\delta_{a}$ are positive (their
magnitudes are always smaller than $\eta_{\omega}$), the $T_c$
reduction would be simply in between the case of a s-wave (no
suppression) and the case of a d-wave. But in the $\pm$s-wave case
$\delta_{a}$ will have always opposite signs and as a result the
$T_c$ reduction can be faster or slower than the d-wave case of
Eq.(15). A simple rule is the following: in the leading
approximation the reduction rate depends on the sign of the
quantity $(\delta_h + \delta_e)$.  If it is positive, the $T_c$
reduction is slower than the d-wave case, and if it is negative,
the $T_c$ reduction is faster than the d-wave case.

We can utilize the relation $|\Delta_h|/|\Delta_e| = \sqrt{
N_e/N_h}$ as $T \rightarrow T_c$ found in the minimal two band
model in Ref.\cite{Bang08}, and obtain
\begin{eqnarray}
\delta_h &\approx& \sqrt{\tilde{N_h}} (\sqrt{\tilde{N_h}}-
\sqrt{\tilde{N_e}}),\\
\delta_e &\approx& - \sqrt{\tilde{N_e}} (\sqrt{\tilde{N_h}}-
\sqrt{\tilde{N_e}}).
\end{eqnarray}
From this we can find that $(\delta_h + \delta_e) \approx
(\sqrt{\tilde{N_h}}- \sqrt{\tilde{N_e}})^2 $ is always positive
regardless whether $N_h > N_e$ or $N_h < N_e$.
Therefore, the actual $T_c$ reduction should be slower than the
d-wave case. How much slower will be determined by the magnitude
of $(\delta_h + \delta_e)$ compared to 1 (s-wave limit) and 0
(d-wave limit). From the relation $(\delta_h + \delta_e) \approx
(\sqrt{\tilde{N_h}}- \sqrt{\tilde{N_e}})^2 $ we can guess that the
the $T_c$ suppression rate is rather close to the d-wave case
because the quantity $(\delta_h + \delta_e)$ is $\ll 1$ unless the
difference of the DOSs between the bands is unrealistically large.
In reality, there are more than two bands and also the intraband
interactions -- which was neglected in the above analysis -- would
make a simple analysis rather difficult. However, a practical rule
of thumb is that the $T_c$ suppression rate by non-magnetic
impurities in the $\pm$s-wave state should be quite similar to the
$d$-wave case. We will show the numerical results obtained by
directly solving the gap Eqs.\ (1) and (2) below.

{\it Magnetic Impurity Case --} Let us now turn to the magnetic
impurity scattering case. For magnetic impurities, if we assume
only an exchange coupling such as ${\bf S} \cdot \vec{\sigma}$
(where ${\bf S}$ is the momentum of the impurity atom and
$\vec{\sigma}$ is the spin of the electrons), we can draw a simple
result from the above analysis. Because the exchange coupling
flips the spin part of the singlet wave function \cite{AG-imp},
the result of the magnetic impurity scattering is to change the
sign of $\Delta_a$ in the numerator of Eq.(11). The final result
in the reduced gap equation is to replace $\delta_a$ by
$-\delta_a$ but keeping $\eta_{\omega}$ the same in Eq.(13). For
the non-magnetic impurities for the $\pm$s-wave pairing, we had
the relation
 \ba
 \label{largerthan1}
(1+\delta_e)(1+\delta_h)>1.
 \ea
It is equal to 1 for a $d$-wave pairing state. Eq.\
(\ref{largerthan1}) was the very reason why the non-magnetic
impurity suppression rate in the $\pm$s-wave state is slower than
the $d$-wave case.

Now, for the magnetic impurity scattering, we have, as discussed
above
 \ba
 \label{smallerthan1}
(1-\delta_e)(1-\delta_h)<1.
 \ea
It is then immediately clear that the magnetic impurity
suppression rate of $T_c$ for $\pm$s-wave pairing should be faster
than the $d$-wave case. For realistic parameter values, the
suppression rates are, as shown in Figs.\ 1 and 2 from numerical
calculations, only marginally faster than the $d$-wave state. In
Fig.\ 3 we used exaggerated parameter values to demonstrate this
the point more clearly. This result, however, is only of an
academic interest because most of magnetic impurity atoms would
have a much larger potential interaction than the exchange
interaction.

\begin{figure}
\noindent
\includegraphics[width=90mm]{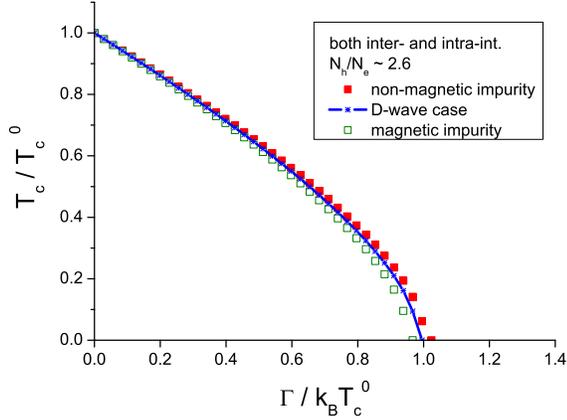}
\caption{(Color online) Normalized critical temperature $T_c /T_c
^0$ vs normalized impurity scattering strength $\Gamma/k_B T_c ^0$
($c=0$). The calculations are with the realistic bands $N_h /N_e
\approx 2.6$ and with the full interactions $V_{hh}, V_{ee},
V_{he}$ and $V_{eh}$. \label{fig1}}
\end{figure}

\begin{figure}
\noindent
\includegraphics[width=90mm]{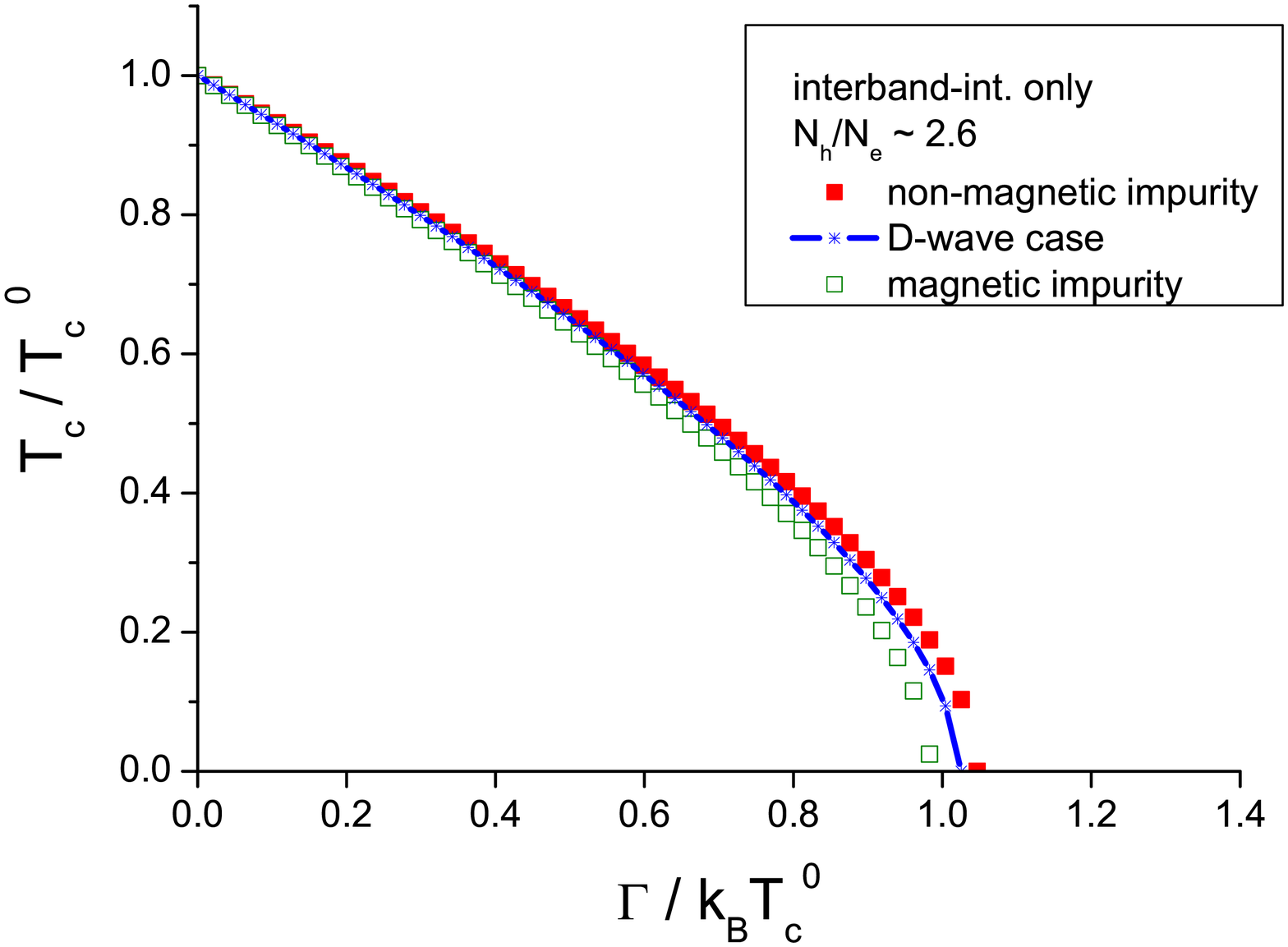}
\caption{(Color online) Normalized critical temperature $T_c /T_c
^0$ vs normalized impurity scattering strength $\Gamma/k_B T_c ^0$
($c=0$). The calculations are with the realistic bands $N_h /N_e
\approx 2.6$ and with the interband interactions  $ V_{he}$ and
$V_{eh}$ only. \label{fig2}}
\end{figure}

\begin{figure}
\noindent
\includegraphics[width=90mm]{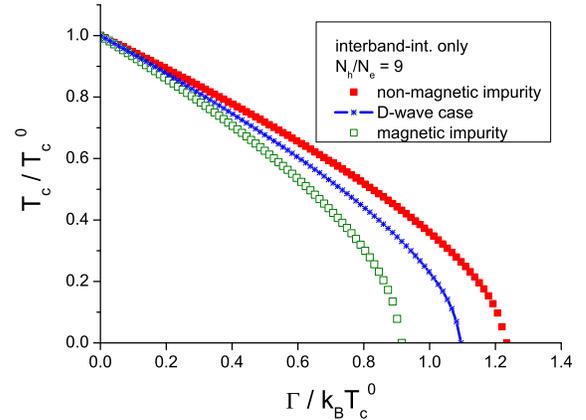}
\caption{(Color online) Normalized critical temperature $T_c /T_c
^0$ vs  normalized impurity scattering strength $\Gamma/k_B T_c
^0$ ($c=0$). The calculations are with the artificial bands $N_h
/N_e = 9.0$ and with the interband interactions $ V_{he}$ and
$V_{eh}$ only. \label{fig3}}
\end{figure}

{\it Numerical Results --} With the typical band structure of the
Fe-based pnictides \cite{band}, we obtained $N_h (0)/N_e (0)
\approx 2.6$ in the previous calculations of Ref.\cite{Bang08}.
With this realistic parameter and all interactions included, Fig.1
shows the calculation results of normalized critical temperatures
$T_c /T_c ^0$ vs normalized impurity scattering strength
$\Gamma/k_B T_c ^0$ in the unitary limit scattering ($c=0$).
Weaker limit of impurity scattering, for example, with $c=1$ would
just yields twice slower suppression rate. As can be seen, there
are almost no differences among all three cases. This is
consistent with our analytic analysis because $(\delta_h +
\delta_e) \approx 0.104 \ll 1$ in this case. Note that the
normalization of the impurity scattering strength $\Gamma$ by $T_c
^0$ instead of using the gap values $\Delta_{h,e}$ at $T=0$ is for
convenience for comparison with future experiments. Also the fact
that $\Gamma/k_B T_c ^0 \sim 1$  when $T_c /T_c ^0 \rightarrow 0$
is a pure coincidence of the parameter choice, which is clear in
Fig.3.

In Fig.2, we artificially shut down the intraband interactions
$V_{hh}$ and $V_{ee}$; without the intraband repulsions $T_c ^0$
itself increases by about 40$\%$.
But the normalized $T_c / T_c ^0$ vs the impurity scattering
strength $\Gamma/k_B T_c ^0$ are indistinguishably the same as the
case of Fig.1. This result shows that our main analytic analysis
for the $T_c$ suppression with the interband interaction only will
be valid for more complicate multiband model in general.

In Fig.3, we artificially increase the DOS ratio to $N_h /N_e = 9$
which is of course unrealistic ratio; this unrealistic parameter
yields $(\delta_h + \delta_e) \approx 0.4$. The calculations
demonstrate that the suppression rate of $T_c$ indeed follow the
trend that we found from the analytic estimation. It also
demonstrates that in realistic case the $T_c$ suppression of the
$\pm$s-wave state by either magnetic or non-magnetic impurities
should be indistinguishably close to the case of the standard
d-wave SC.

{\it Conclusions --} In summary, we studied the effect of
impurities for the $T_c$ suppression on the $\pm$s-wave SC using a
generalized $\mathcal{T}$-matrix method. The main finding is that
{\it despite a possibly large difference of the positive and
negative s-wave OP magnitudes, the $T_c$ suppression rate is
practically indistinguishable from the standard d-wave case.} This
is because the DOS enters together with the OP for the scattering
process. As a by-product, we found the subtle difference between
the magnetic and non-magnetic impurities for the $T_c$
suppression, which should, however, be a quite small difference in
realistic case.

{\it Acknowledgement -- } The author (YB) acknowledges useful
discussion  with O. Dolgov.  This work was supported by the KOSEF
through the Grants No. KRF-2007-521-C00081 (YB),and No.
KRF-2007-070-C00044 (YB,HYC), and Basic Research Program Grant No.
R01-2006-000-11248-0 (HYC).


\begin{references}


\bibitem{YKamihara} Y. Kamihara {\it et al.}, J. Am. Chem. Soc. {\bf 128}, 10012 (2006);
Y. Kamihara, T. Watanabe, M. Hirano, and H.
Hosono, J. Am. Chem. Soc. {\bf 130}, 3296 (2008).

\bibitem{PhysToday08} Phys. Today {\bf 61}, Issue 5, 11 (2008); G.
F. Chen et al., Phys. Rev. Lett. {\bf 100}, 247002 (2008); G. F.
Chen et al., Nature {\bf 453}, 761 (2008).

\bibitem{Mazin08}
 I.I. Mazin, D.J. Singh, M.D. Johannes, M.H. Du, Phys. Rev. Lett. {\bf 101},
057003 (2008).

\bibitem{Choi08} H.-Y. Choi and Y. Bang, arXiv:0807.4604.

\bibitem{Suhl59} H. Suhl, B. T. Matthias, and L. R. Walker, Phys.
Rev. Lett. {\bf 3}, 552 (1959).

\bibitem{Rice91} M. J. Rice, H. Y. Choi, and Y. R. Wang, Phys.
Rev. B {\bf 44}, 10414 (1991).

\bibitem{pene}
 L. Malone el al., arXiv:0806.3908 (unpublished);  K. Hashimoto et al.,
Phys. Rev. Lett. {\bf 102}, 017002 (2009);  C. Martin et al.,
arXiv:0807.0876 (unpublished).

\bibitem{T1}
K. Matano et al., Europhys. Lett. {\bf 83}  57001 (2008); H.-J.
Grafe et al., Phys. Rev. Lett. {\bf 101}, 047003 (2008); H. Mukuda
et al., J. Phys. Soc. Jpn. {\bf 77} (2008) 093704; Y. Nakai et
al., J. Phys. Soc. Jpn. {\bf 77} (2008) 073701.

\bibitem{Bang08b} Y. Bang and H.-Y. Choi, Phys. Rev. B {\bf 79}, 054529 (2009).

\bibitem{recent}
D. Parker, O.V. Dolgov, M.M. Korshunov, A.A. Golubov, I.I. Mazin ,
Phys. Rev. B {\bf 78}, 134524 (2008); A.V. Chubukov, D. Efremov,
I. Eremin, Phys. Rev. B {\bf 78}, 134512 (2008); M. M. Parish, J.
Hu, B. A. Bernevig, Phys. Rev. B {\bf 78}, 134512 (2008).


\bibitem{Bang08} Y. Bang and H.-Y. Choi, Phys. Rev. B {\bf 78}, 134523 (2008).

\bibitem{AG-imp}
A.A. Abrikosov and L.P. Gorkov, Sov. Phys. JETP, {\bf 12}, 1243
(1961).


\bibitem{band}
D.J. Singh and M.-H. Du, Phys. Rev. Lett. {\bf 100}, 237003 (2008)
; C. Cao, P. J. Hirschfeld, H. Cheng, Phys. Rev. B {\bf 77},
220506 (2008); E. Manousakis, Jun Ren, E. Kaxiras, Phys. Rev. B
78, 205112 (2008).



\end{references}
\end{document}